# Neutralization of the impact of belt speed on screen printed copper metallization by LECO on PERC homogeneous emitter

Abasifreke Ebong[1], Donald Intal[1], Sandra Huneycutt[1], Ajeet Rohatgi[2], Vijay Upadhyaya[2], Sagnik Dasgupta[2], Ruohan Zhong[2], Thad Druffel[3], Ruvini Dharmadasa[3]

[1]University of North Carolina at Charlotte, Charlotte, NC 28223, USA

[2]Georgia Institute of Technology, Atlanta, Georgia 30332, USA

[3]Bert Thin Films LLC, Louisville, KY 40208, USA

*Abstract*— **Copper fire-through metallization is a cost-effective alternative to Ag counterpart for industrial high efficiency solar cells. The fire through dielectric metallization relies on belt speed, which dictates the ramp up and ramp down rates for effective contact formation. In this paper three belt speeds (325°C, 360°C, 390°C) at constant peak firing temperature, were used to process PERC (homogeneous emitter) cells. After the contact firing the electrical parameters were dependent on belt speed, but after LECO treatment, they were identical. The SEM/EDS cross sectional analyses showed increased elemental Cu with belt speed, and the series resistance was lowest for the middle belt speed before LECO. However, after the LECO treatment, the series resistance dropped, respectively, to 0.503 Ωcm$^{-2}$, 0.428 Ωcm$^{-2}$ and 0.500 Ωcm$^{-2}$ leading to efficiency of 20.8% on homogeneous PERC emitter.**

## I. Introduction

Silicon solar cells remain the dominant photovoltaic technology because they combine a mature manufacturing ecosystem with continued opportunities for efficiency improvement through passivation and contact engineering. In this setting, front-side metallization is both a cost driver and a performance limiter [1]. It must deliver low resistive loss and stable adhesion while minimizing optical shading and recombination at the metal silicon interface. Silver pastes have long enabled reliable, high throughput screen printing, but cost pressure and supply volatility continue to motivate alternatives that can be deployed at scale.

Copper is a compelling candidate because it offers lower material cost while maintaining high electrical conductivity [2]. At the same time, Cu based metallization is often more sensitive to contact formation conditions than conventional Ag processing, which can narrow the practical range of conditions in which low resistive loss and low leakage are achieved simultaneously [1,3]. This sensitivity is especially important in device designs that rely on high quality surface passivation and shallow emitters, where small changes in contact formation can strongly influence voltage, fill factor, and long-term stability. As a result, understanding how screen-printed fire-through Cu contacts form and how their microstructure evolves is essential for implementing Cu metallization without sacrificing performance or reliability [4].

In screen-printed fire-through systems, contact behavior is governed by the microstructure created during firing [1]. Sintering and related thermally activated transformations govern the coupled steps that translate a printed Cu paste into an electrically functional contact [5–7]. Firing drives Cu particle coalescence and densification within the printed gridline and influences how current is transferred through the dielectric stack into the emitter, while also affecting the likelihood of contact related leakage pathways. The outcome is a contact microstructure that can support uniform current extraction when the Cu network is continuous, and the interface is well distributed or can instead promote resistive loss and non-uniform current flow when the metallization remains porous or electrically discontinuous. Conversely, overly localized interfacial activation can increase leakage susceptibility and reduce shunt resistance and fill factor [1,8].

Despite the central role of microstructure, many reports emphasize device level metrics without a coordinated set of diagnostics that connect contact morphology, local current transport, and wafer scale non uniformity for Cu fire-through contacts [3,9–12]. In this work, we quantify firing-condition sensitivity in screen-printed fire-through Cu metallization on industrial silicon solar cells and evaluate how laser-enhanced contact optimization (LECO) modifies this sensitivity. Belt-speed variation is used to modulate high-temperature dwell time, and performance is assessed using illuminated IV analysis together with low-injection diagnostics to distinguish transport-limited behavior from parasitic-loss contributions. Cross-sectional SEM/EDS is used to relate these electrical signatures to contact morphology and interfacial composition, establishing a direct linkage between firing-driven microstructural evolution and device operating regimes.

## II. Experimental Method

G1-format industrial crystalline Si solar cells were fabricated on boron-doped p-type Si with a phosphorus-diffused $n^+$ front emitter and a passivated front surface. Front-side current collection employed screen-printed, fire-through Cu busbars and gridlines. Contacts were fired in an industrial inline belt furnace (TPSolar) using a fixed firing profile with a peak temperature of approximately 560–600 °C. Three dwell-time conditions were established by setting belt speed to 325, 360, and 390 inches per minute (IPM) (BS325, BS360, and BS390).

Illuminated IV measurements under standard test conditions were performed using a Sinton FCT-450 to determine $V_{oc}$, $J_{sc}$, fill factor, and efficiency; series resistance ($R_s$) and shunt resistance ($R_{sh}$) were derived from the IV curves. To relate firing condition to contact behavior, cross-sectional FE-SEM (Hitachi SU8700) with EDS was used to evaluate contact morphology and interfacial composition, and electroluminescence imaging was used to assess wafer-scale nonuniformities associated with resistive loss and shunting.

## III. Results and Discussion

*Table 1: IV parameters across three belt speeds for pre- and post-leco sample.*

| Params | Belt Speed (IPM) | | | | | |
|---|---|---|---|---|---|---|
| | 325 | | 360 | | 390 | |
| | Pre-LECO | Post-LECO | Pre-LECO | Post-LECO | Pre-LECO | Post-LECO |
| $V_{OC}$ (V) | 0.663 | 0.662 | 0.648 | 0.662 | 0.647 | 0.663 |
| FF (%) | 51.65 | 80.13 | 56.29 | 80.41 | 52.91 | 80.12 |
| Eff. (%) | 13.40 | 20.78 | 14.17 | 20.79 | 13.25 | 20.78 |
| $R_{SH}$ (Ω-cm²) | 101524 | 100543 | 15074 | 77876 | 34180 | 98858 |
| $R_S$ (Ω-cm²) | 6.252 | 0.503 | 4.561 | 0.428 | 5.406 | 0.500 |
| $n_{1sun}$ | 0.991 | 0.992 | 0.989 | 0.984 | 1.007 | 0.987 |
| $n_{1/10sun}$ | 1.158 | 1.166 | 1.510 | 1.199 | 1.624 | 1.177 |
| pFF (%) | 83.02 | 82.90 | 81.34 | 82.75 | 80.86 | 82.86 |
| $J_{02}$ (A/cm2) | 6.50E-09 | 7.21E-09 | 3.01E-08 | 8.98E-09 | 3.64E-08 | 7.83E-09 |
| $J_{SC}$ (mA/cm2) | 39.11 | 39.14 | 38.39 | 39.03 | 38.21 | 39.09 |

Across the investigated firing window, the Cu fire-through metallized cells show a clear belt-speed dependence prior to Laser-Enhanced Contact Optimization (LECO), followed by a strong post-LECO convergence. Although the firing profiles are similar in peak temperature and cooling rate (Figure 1), belt speed modulates the effective high-temperature dwell time and, correspondingly, the as-fired electrical behavior [6,7]. Before LECO, Voc varies only modestly (0.647–0.663 V) and Jsc remains near 39 mA·cm⁻² (Table 1), indicating that belt-speed-dependent differences are not primarily governed by optical collection or bulk-limited recombination. Instead, the performance spread is dominated by changes in FF and efficiency, while pFF remains comparatively high, implying that intrinsic diode quality is largely retained even when operating-point performance is not [1].

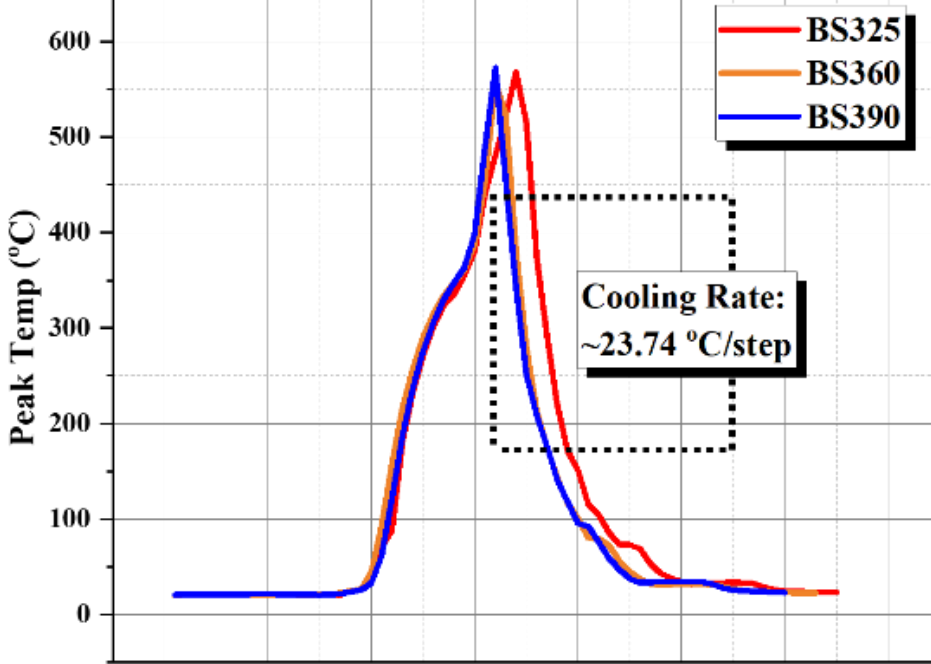

*Figure 1: Firing profile across three belt speeds.*

Figure 2 shows the persistent separation between FF and pFF points to resistive and transport-related losses that prevent the intrinsic diode response from being realized underload. Consistent with this interpretation, efficiency tracks FF across belt speed despite the near invariance of Voc and $J_{sc}$. All pre-LECO samples exhibit elevated series resistance (Rs = 4.56–6.25 Ω·cm²), consistent with transport-limited operation. Within this window, BS360 provides the lowest $R_s$ and highest FF, yielding the best pre-LECO efficiency and indicating the most effective baseline current transfer between the Cu grid and the silicon emitter. By contrast, BS325 exhibits the highest $R_s$ and lower efficiency, consistent with insufficient contact formation at lower effective thermal budget. Increasing belt speed beyond this intermediate condition does not improve performance; BS390 shows reduced FF and efficiency without a commensurate reduction in $R_s$, suggesting additional loss pathways beyond series resistance.

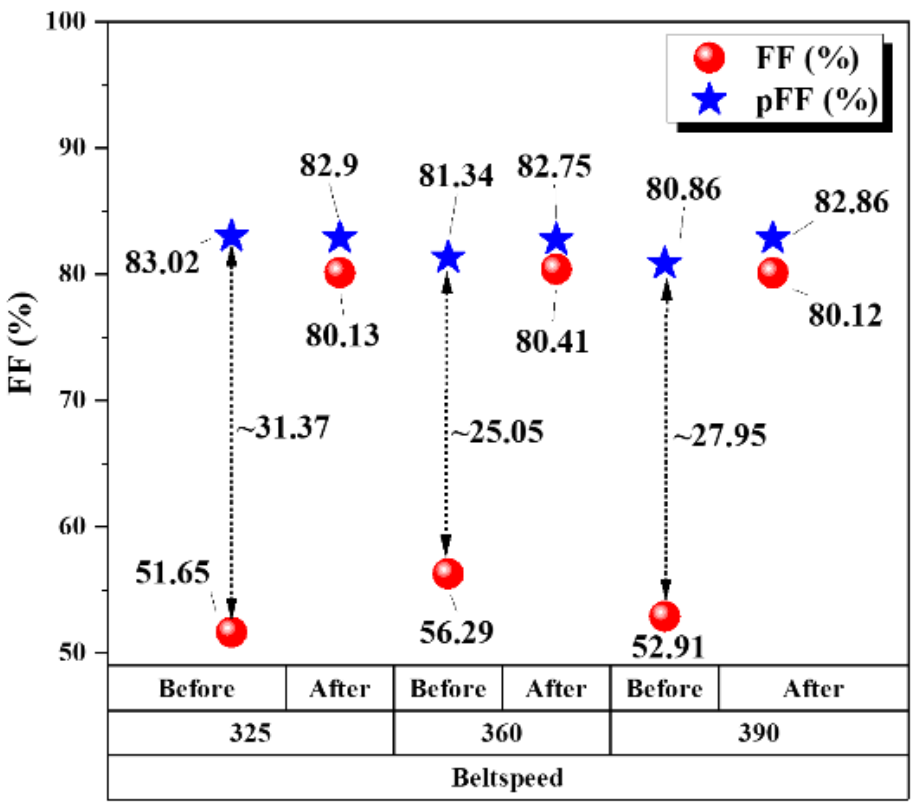

*Figure 2: Comparison of fill factor, and pseudo fill factor as a function of firing belt speed for pre- and post-leco sample.*

Those additional pathways are most evident under low injection. As illustrated in Figure 3, shows that prior to LECO, BS390 shows a pronounced increase in ideality factor at 0.1 sun (n@0.1sun = 1.62) compared with BS325 (1.20) and BS360 (1.16), accompanied by reduced $R_{sh}$ and an increased recombination-current component $J_{02}$ (Figure 3). This behavior is inconsistent with a uniform shift in bulk recombination and instead indicates parasitic current components that become dominant under weak illumination. The absence of a large $V_{oc}$ collapse supports a localized (spatially heterogeneous) origin rather than a globally distributed degradation mechanism.

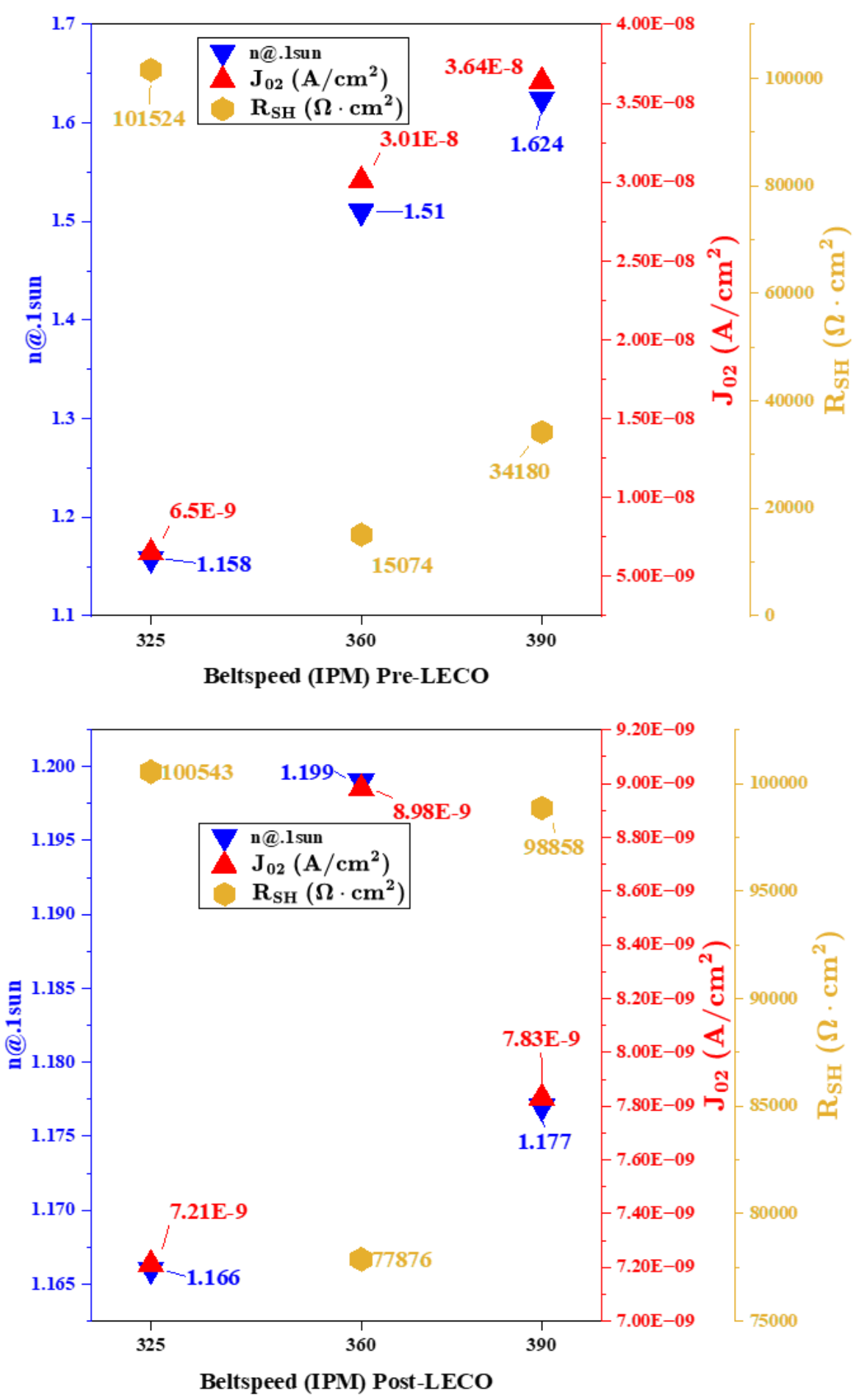

*Figure 3: Extracted Ideality factor, shunt resistance (Rsh), and Jo2 as a function of firing belt speed for pre- and post-leco sample.*

The belt-speed sensitivity of the low-injection diagnostics implicates firing-induced changes near the metal–silicon interface, where current is injected into the emitter [1]. As shown in Figure 4, SEM/EDS collected along the Cu–Si interface shows an increased prevalence of Cu-rich particulate features with increasing belt speed, suggesting retention of discrete Cu-rich regions under shorter effective thermal exposure. However, the electrical response demonstrates that Cu presence alone is not sufficient to ensure effective current transfer: in BS390, a higher prevalence of Cu-rich interfacial features coexists with degraded low-injection behavior, indicating an interface that is chemically Cu-rich yet electrically heterogeneous. Such heterogeneity is particularly detrimental for a homogeneous, high-sheet-resistance emitter, where current injections through a limited subset of functional contact sites increases lateral transport burden and drives current crowding [13,14]. Elevated local current density and junction-adjacent electric fields provide a consistent link between BS390 and the observed increases in n@0.1sun and $J_{02}$, reductions in $R_{sh}$, and the large FF–pFF separation characteristic of transport-limited operation.

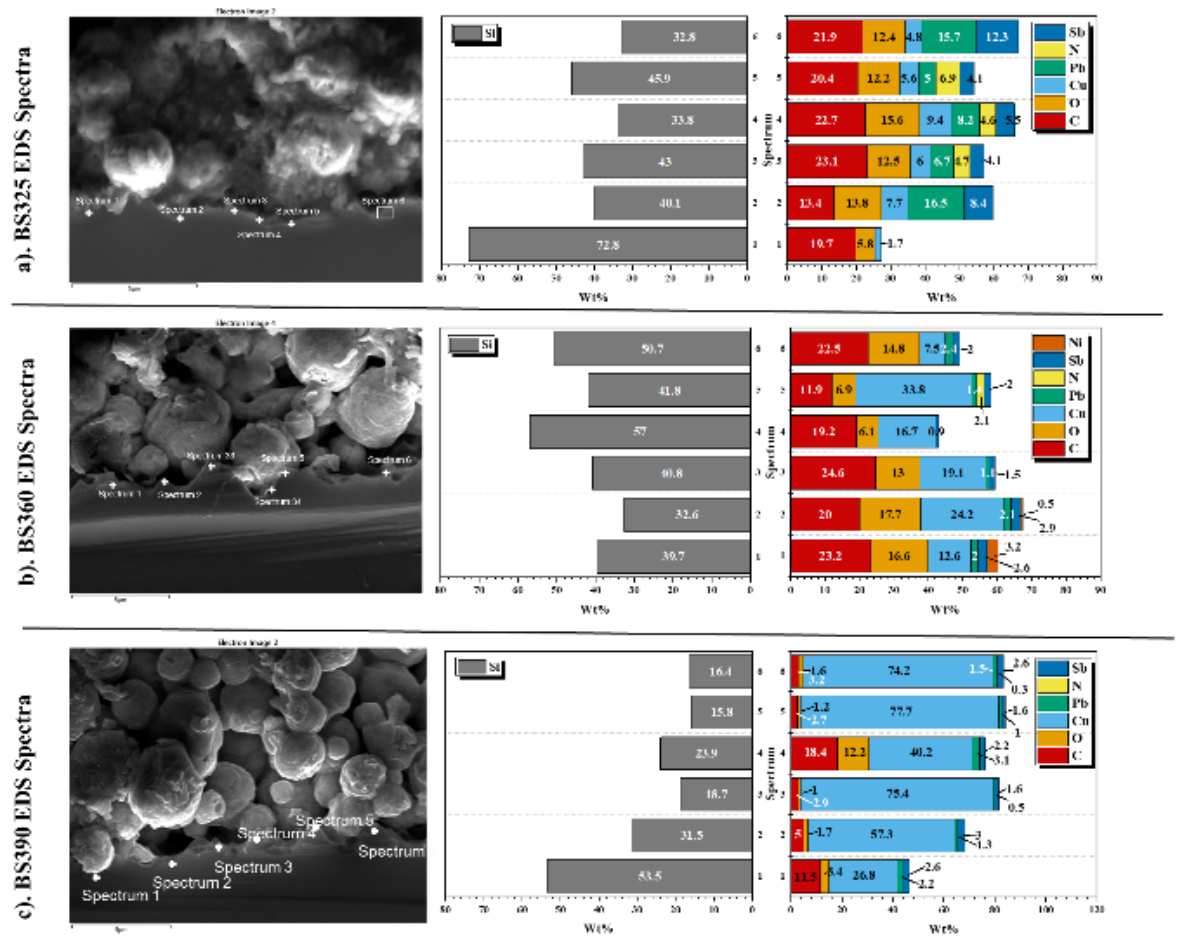

*Figure 4: Energy-dispersive X-ray spectroscopy (EDS) analysis of the Cu/Si interfacial region for the three samples.*

After LECO, the belt-speed dependence is largely removed, and performance converges across all conditions. Rs contracts to 0.43–0.50 Ω·cm², FF rises to ~80%, and the FF–pFF separation narrows to ~2–3 percentage points, accompanied by suppression of the low-injection degradation most strongly for the most degraded condition. Overall, the coupled recovery of operating-point performance ($R_s$, FF) and low-injection diagnostics (n@0.1sun, $J_{02}$, $R_{sh}$) indicates that LECO mitigates firing-induced non-uniformity at the metal–emitter interface. By increasing the fraction and spatial uniformity of electrically effective injection sites, LECO reduces current crowding and associated parasitic pathways, allowing the intrinsically high diode quality (high pFF) to translate into operating-point performance across the investigated belt-speed window.

# IV. Conclusions

In this work, we show that even though the SEM/EDS analysis depicted increased elemental Cu with belt speed, the efficiency after LECO for the three belt speeds is the same. Thus, LECO neutralizes the series resistance difference and fill factor that existed between the belt speeds and harmonizes the efficiency to 20.8% on homogeneous PERC emitter. The effectiveness of LECO largely removes this belt-speed sensitivity, producing convergence in Rs, FF, and low-injection metrics and enabling the intrinsically high diode quality (high pFF) to be realized under operating conditions. These results indicate that LECO is an effective route to mitigate firing-induced interfacial non-uniformity and broaden the viable firing process window for Cu fire-through metallization.